\documentclass[12pt]{article}
\pdfoutput=1
\usepackage{amsmath,amssymb}

\usepackage{graphicx}
\usepackage{color}
\usepackage[bookmarksnumbered=true,colorlinks=true,linkcolor=black,citecolor=black]{hyperref}
\newlength{\extraspace}
\newlength{\extraspaces}
\def\bsklength{.8mm} 

\newcommand{\beq}{\begin{equation}}
\newcommand{\eeq}{\end{equation}}

\newcommand{\bseq}{\addtocounter{subeqno}{1}\begin{subequations}}
\newcommand{\eseq}{\end{subequations}}

\newcommand{\newsection}[1]{
\vspace{6mm}
\pagebreak[3]
\addtocounter{section}{1}
\setcounter{subsection}{0}
\setcounter{figure}{0}
\phantomsection%
\addcontentsline{toc}{section}{\protect\numberline{\arabic{section}.}{#1}}
\noindent{\large \bf \thesection. #1}
\nopagebreak
\medskip
\nopagebreak}

\font\mathscript=eusm10 at 12pt
\font\mathscripts=eusm7
\font\mathscriptss=eusm5
\newfam\mathscri
\textfont\mathscri=\mathscript
\scriptfont\mathscri=\mathscripts
\scriptscriptfont\mathscri=\mathscriptss
\def\mathscr#1{{\fam\mathscri\relax#1}}

\font\mathfrakt=eufm10 at 12pt
\font\mathfrakts=eufm7
\font\mathfraktss=eufm5
\newfam\mathfraki
\textfont\mathfraki=\mathfrakt
\scriptfont\mathfraki=\mathfrakts
\scriptscriptfont\mathfraki=\mathfraktss
\def\mathfrak#1{{\fam\mathfraki\relax#1}}

\def\CO{{\cal O}}

\renewcommand{\bar}{\overline}
\def\half{{\textstyle{1\over 2}}}

\newcommand{\e}{{\rm e}}

\begin{document}
\setcounter{page}{0}
\addtolength{\baselineskip}{\bsklength}
\thispagestyle{empty}
\renewcommand{\thefootnote}{\fnsymbol{footnote}}        

\begin{flushright}
\end{flushright}
\vspace{.4cm}

\begin{center}
{\Large
{\bf{Ekpyrotic Reheating and Fate of Inflaton}}}\\[1.2cm]
{\rm HoSeong La\footnote{hsla.avt@gmail.com}
}
\\[3mm]
{\it Department of Physics and Astronomy,\\[1mm]
Vanderbilt University,\\[1mm]              
Nashville, TN 37235, USA} \\[1.5cm]

\vfill
{\parbox{15cm}{
\addtolength{\baselineskip}{\bsklength}
\noindent
It is shown that perturbative reheating can reach a
sufficiently high temperature with small or negligible inflaton decay rate
provided that the inflaton potential becomes negative after inflation.
In our model, inflaton and dark energy field are two independent scalar
fields, and, depending on the mass of the inflaton and its coupling to 
matter fields, there is a possibility that the remaining inflaton after 
reheating can become a dark matter candidate.

}
}


\end{center}
\noindent
\vfill


\newpage
\setcounter{page}{1}
\setcounter{section}{0}
\setcounter{equation}{0}
\setcounter{footnote}{0}
\renewcommand{\thefootnote}{\arabic{footnote}}  
\newcounter{subeqno}
\setcounter{subeqno}{0}
\setlength{\parskip}{2mm}
\addtolength{\baselineskip}{\bsklength}

\pagenumbering{arabic}


\newsection{Introduction}

Modern cosmology\cite{KT,Dodelson:2003ft} based on the inflation 
demands three seemingly different 
types of particles, whose identities are still elusive.
Inflaton that drives inflation is one, dark matter (DM) is another, and
the third one is to explain the nature of dark energy. There have been 
attempts to unify these different fields for the sake of 
simplicity\cite{Liddle:2006qz,PerezLorenzana:2007qv,Liddle:2008bm}.
Some have tried to reduce the number of fields by relating some of them.
For example, ref.\cite{Peebles:1998qn} considers a connection between 
the inflaton and dark energy. In this paper, we propose a possibility to link
the inflaton and DM\footnote{See 
\cite{Allahverdi:2007wt}\cite{delaMacorra:2012sb} 
for different proposals.}, 
while introducing a new mechanism of reheating. 

This paper is organized as follows. In section 2, the basic structure of 
the model is described. In section 3, it is shown that the perturbative 
reheating can reach high enough temperature, which imitates 
the ``ekpyrotic" phase. In section 4, since the entire disappearance of 
the inflaton may not be needed for the sufficient reheating, the possibility
of remaining inflaton as a dark matter candidate is explained.
Finally, in section 5, some discussions are provided.

\newsection{The Model}

The model we consider has two scalar fields: one is the inflaton that 
drives inflation, while the other is responsible for later expansion of 
the universe as dark energy. The total (effective) potential energy takes 
the form of
\beq
V(\phi,\eta) = V_1(\phi) + V_2(\eta)\geq 0,
\eeq
where $\phi$ is the inflaton and $\eta$ is the (late time)
dark energy field (DEF). $\phi$ carries no Standard Model charges and 
interacts with fermions only in terms of Yukawa couplings of 
$\lambda_f\phi\bar{f}f$, while $\eta$ does not interact even with fermions.
We assume that $V_2(\eta)$ is invariant under $\eta\to\eta + a$,
where $a$ is an arbitrary constant.
Note that $\phi$ and $\eta$ do not interact with each other.
An important assumption is that $V_1(\phi)$ becomes negative 
at some point\footnote{A negative scalar potential is also considered in 
\cite{hep-th/0307132,arXiv:1106.1416}, in which the negative part of the potential is identified as big crunch before the big bang. 
In our case, however, there is no crunch because of $\eta$
such that $V(\phi,\eta)\geq 0$ always.}, 
and $V_2(\eta)$ becomes dominant at later time 
to sustain the expansion of the universe.
We also assume that $\eta$ remains massless to avoid any localization of 
dark energy, while inflaton $\phi$ could become massive.
Since we could actually take $V_2(\eta)$ to be
constant\cite{Liddle:2006qz,Liddle:2008bm} for simplicity,
although it can be more complicated, we choose
\beq
V_2(\eta) = -V_1(\phi_{\rm min})
\eeq
such that
\beq
V(\phi,\eta) 
\begin{cases} 
=0  &{\rm if}\ \phi=\phi_{\rm min},\\
> 0 &\mbox{otherwise}.
\end{cases}
\eeq
This choice can actually avoid fine-tuning because the magnitude of the 
current cosmological constant becomes equivalent to the vacuum fluctuation 
of the inflaton around the true vacuum, which can be small enough. 

Note that in quantum theories the potential energy cannot be shifted up or down 
arbitrarily because the minimum value (of the effective potential) 
has a definite physical meaning, unlike 
in some classical theories, where only relative values have physical meaning. 
For decoupled fields, in particular, their ground state values can be in 
principle separately measurable.
Since the inflaton and DEF are two independent fields without a direct 
coupling, it is legitimate to assume their separate ground state energy levels
and even constant parts of their potential energies should not be 
exchanged.
If a different minimum value is chosen for the inflaton potential, 
the equation of state changes so that it becomes a different physical system.

\begin{figure}[t] 
\begin{minipage}[b]{0.5\linewidth}
\centering
\includegraphics[width=2.8in]{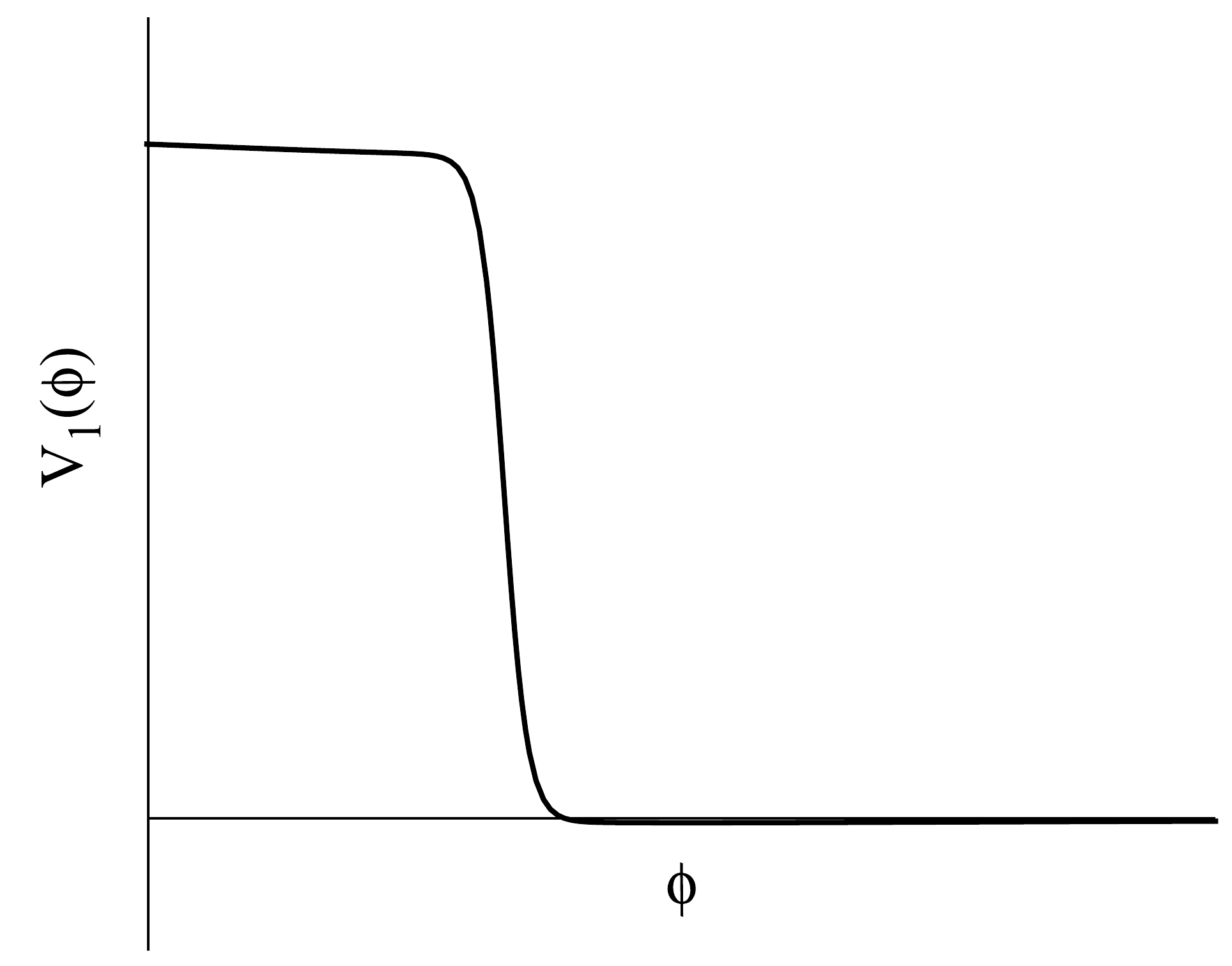} 
   \parbox{0.9\textwidth}{\vskip-6pt
   \caption{Inflaton potential for $b\ll 1$ in eq.(\ref{e:vp1}).}
   \label{fig:2}}
\end{minipage}
\hspace{0.1cm}
\begin{minipage}[b]{0.5\linewidth}
\centering
   \includegraphics[width=2.8in]{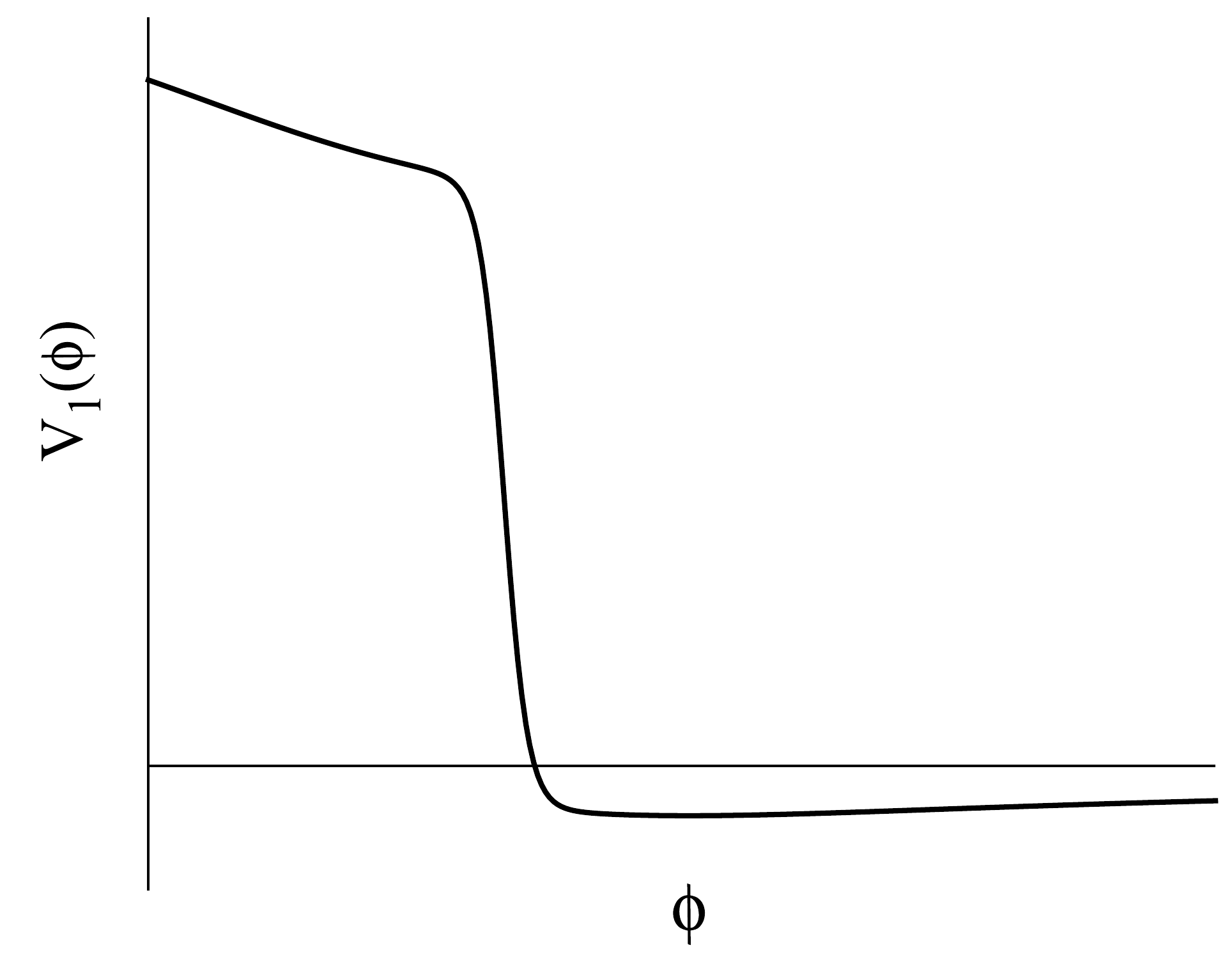} 
   \parbox{0.9\textwidth}{\vskip-6pt
   \caption{Exaggerated inflaton potential for larger $b$ 
   in eq.(\ref{e:vp1}).}
   \label{fig:3}}
\end{minipage}
\end{figure}

The ideal (effective) potential for the inflaton could be visualized 
numerically, for example, as
\beq
\label{e:vp1}
V_1(\phi) = {1\over \e^{a(\phi-\phi_0)}+1}+b\left(1-\e^{-c\phi^2}\right)
\left({1\over \phi^2}-{d\over \phi}\right),
\eeq
where $a,b,c,d,\phi_0$ are constants, and shown in Fig.\ref{fig:2} 
and Fig.\ref{fig:3}. 
This provides a slow-roll potential initially, then becomes steep to change 
to negative toward 
a stable vacuum, and, finally, approaches zero asymptotically.
The parameter $\phi_0$ controls the number of e-foldings $N_e$.
What we further need is $\phi_{\rm min}$ should be close to $\phi_0$.
Note that, in arbitrary units, for $b\ll 1$, i.e. assuming $b$ is small 
so that the second term is negligible in computing $N_e$, then
\beq
\phi_0 \simeq {1\over a}\ln\left({N_e a^2\over 8\pi}\right).
\eeq
To get the needed e-foldings $N_e\simeq 60$, for example, for $a=40$, we need
$\phi_0\simeq 0.21$. To get a reasonable $\phi_{\rm min}$
close enough to $\phi_0$, we need to adjust $c, d$ properly.
The natural unit expression can be obtained by assigning a suitable energy 
scale to $\phi_0$, etc., but it is not essential for our purpose.

\newsection{``Ekpyrotic" Reheating}

The universe cools off during inflation, so it needs to reheat
to a sufficiently high temperature in the radiation-dominated era\cite{KT}. 
(See \cite{arXiv:1001.2600} for a recent review.)
It is commonly known that the usual perturbative reheating cannot 
generate a sufficient temperature due to normally small coupling constant 
between the inflaton and matter fields. 
To overcome this, the idea of preheating is introduced in 
\cite{Traschen:1990sw,KLS}, but it requires a coupling between the 
inflaton and another scalar field.\footnote{See \cite{Greene:2000ew} 
for a proposal of preheating with fermionic coupling. 
However, preheating cannot take place until the inflaton becomes massive.}
However, if one can achieve enough reheating, the preheating is unnecessary
and an extra scalar field can be avoided.
In this paper, we shall propose a new perturbative reheating mechanism
that can generate a sufficiently high reheating temperature.

A sufficiently high reheating temperature requires rapid transfer of 
the inflaton energy to radiation right after inflation.
There are two ways of achieving this: one is to take advantage of the
coherent nature of the inflaton (near the new vacuum), another is to take 
advantage of rapid growth of inflaton kinetic energy when the slow-roll
potential turns steep\cite{Albrecht:1982mp}. 
Here we shall take the latter path, but it is also related 
to the former implicitly.

Consider the equation of state for the inflaton, 
$p_\phi = w_\phi\rho_\phi$, where
\bseq
\begin{align}
\rho_\phi &= \half\dot{\phi}^2 + V_1(\phi),\\
p_\phi &=\half\dot{\phi}^2 - V_1(\phi),
\end{align}
\eseq
so that the equation of state parameter $w_\phi$ can be expressed as
\beq
\label{e:ep}
w_\phi={\half{\dot\phi}^2-V_1(\phi)\over \half{\dot\phi}^2+V_1(\phi)}.
\eeq
As the inflaton evolves, $w_\phi$ is not constant but varies.
If $V_1(\phi)\geq 0$, 
$1 \geq w_\phi >-1$ always.
However, for $\rho_\phi\geq 0$, if (and only if) $V_1(\phi)<0$, 
$w_\phi>1$ is allowed.
(If $\rho_\phi<0$ for $V_1(\phi)<0$, $w_\phi<-1$.)
Note that, if $V_1(\phi)<0$, $\rho_\phi$ can be as small as possible. 
Hence, as the inflaton loses kinetic 
energy, i.e. $\rho_\phi\to 0$, $w_\phi \gg 1$ can happen.


The possibility of $w_\phi\gg 1$ is the key to successful perturbative reheating
in our model. Hence, we call it {\it ekpyrotic} reheating,
borrowing the word from \cite{arXiv:1106.1416,Khoury:2001wf}. 
In general, the inflaton can lose its energy by decaying into 
radiation, scattering with radiation, and/or annihilating into radiation.
With taking all these processes into account as the source of damping, 
the evolution equation for the inflaton can be generalized for arbitrary
$w_\phi$ as
\beq
\label{e:1}
{\dot{\rho}_\phi\over w_\phi+1} +(3H+\Delta_\phi)\rho_\phi 
+\Sigma_\phi\rho_\phi^2 =0.
\eeq
The prefactor $1/(w_\phi+1)$ is derived from the equation of motion of 
the inflaton and eq.(\ref{e:ep}), with the assumption that $\phi$ is not 
too heavy so that it can be approximately treated as a perfect fluid.
$\Delta_\phi$ in the second term represents the contribution from decay and
scattering processes, and the dominant contribution is from 
the former such that 
\beq
\Delta_\phi =\Gamma_\phi+\cdots,
\eeq
where $\Gamma_\phi$ is the decay rate and the detail of the ellipsis for scattering is not important for our purpose. The third term represents 
the energy loss due to annihilation processes (see eq.(\ref{eP12})) and
\beq
\Sigma_\phi\equiv {\langle\sigma v\rangle\over E_\phi}
\eeq
with some energy scale parameter $E_\phi$. If $\Delta_\phi\neq 0$, the third
term is negligible. However, if $\Delta_\phi = 0$, this should be taken into
account.
If $\phi$ is massive, $E_\phi=m_\phi$. If $\phi$ is massless, the average value
$E_\phi=\rho_\phi(t)/n_\phi(t)\simeq\rho_\phi(t_i)/n_\phi(t_i)\propto T_i$ 
can be taken.
In our model, the inflaton decay rate is solely given by
Yukawa couplings of $\lambda_f\phi\bar{f}f$ as\footnote{We leave the fermion 
content generic because we do not yet know the particle content of light degrees of freedom at the energy scale where the reheating takes place. 
This could be most likely beyond the EW scale.
Nevertheless, to be more specific, for our purpose, any isospin singlet 
fermions with or without U(1) charges, e.g. hypercharges, can satisfy our need.
Such fermions can be easily accommodated in models beyond the SM. For example,
the right-handed neutrino could be one, then the Yukawa coupling is a Majorana 
type. Another example can be isospin singlet Dirac fermions with 
hypercharges without generating anomalies, as long as their masses at the low 
energy are generated beyond the SM energy scale. These fermions subsequently 
annihilate into the SM particles via gravitational or U(1) interactions 
as the Universe evolves.}
\beq
\label{e:pdr}
\Gamma_\phi =\sum_f {\lambda_f^2\over 8\pi}
m_\phi\left(1-4{m_f^2\over m^2_\phi}\right)^{3/2}.
\eeq
Note that this decay process is dominated by the largest Yukawa 
coupling constant
and smallest fermion mass. Typically, if $\phi$ is massive, decay process can 
occur because $\Gamma_\phi\neq 0$. However, if $\phi$ is massless, decay 
cannot happen directly and we have to take into account of other processes.

Note that eq.(\ref{e:1}) is actually valid for time-dependent $w_\phi$.
However, since most of reheating is to take place as the inflaton potential
turns negative and $w_\phi$ at its maximum, where $\phi$ does not vary too much,
$w_\phi$ will be assumed to be constant to solve eq.(\ref{e:1}). 

We shall first demonstrate the ekpyrotic reheating in the case of 
$\Delta_\phi\neq 0$.
Since the annihilation term can be ignored for $\Delta_\phi\neq 0$,
eq.(\ref{e:1}) can be solved to yield
\beq
\label{e:ev2}
\rho_\phi(t) \simeq \rho_\phi(t_i) 
\left({a_i\over a}\right)^{3(w_\phi+1)}
\e^{-(w_\phi+1)\Delta_\phi(t-t_i)},
\eeq
where $a_i\equiv a(t_i)$ and $w_\phi$ is the chosen maximum value. 
Note that $\rho_\phi>0$ and this is an approximate solution since
we assumed $w_\phi$ to be constant.
This shows that $\rho_\phi$ decreases more rapidly if $w_\phi>0$,
compared to the case of $w_\phi=0$.
So, we can already anticipate that this will lead to a higher reheating 
temperature. The energy density of radiation $\rho_R$ satisfies
\bseq
\begin{align}
\label{e:erh1}
\dot{\rho}_R +4H\rho_R 
&= -\dot{\rho}_\phi-3H(1+w_\phi)\rho_\phi -\dot{\rho}_\eta
=(w_\phi+1)\Delta_\phi\rho_\phi-\dot{\rho}_\eta,\\
\label{e:erh2}
H^2 &={8\pi G\over 3}\left(\rho_\phi +\rho_\eta +\rho_R\right),
\end{align}
\eseq
where we assume $\rho_\eta \ll \rho_\phi,\rho_R$ so that we can ignore
$\rho_\eta$ terms in the following.
Since $a(t)\sim t^{2/(3(w+1))}$, where $p_{\rm tot}=w\rho_{\rm tot}$ 
for the universe, such that 
\beq
H\equiv {\dot{a}\over a}={2\over 3(w+1)}{1\over t},
\eeq
eq.(\ref{e:erh1}) with eq.(\ref{e:erh2}) leads to
\beq
0=\dot{\rho}_R +\left\{{8\over 3(w+1)}{1\over t}
+(w_\phi+1)\Delta_\phi\right\}\rho_R 
-{(w_\phi+1)\Delta_\phi\over 6\pi G (w+1)^2}{1\over t^2}.
\eeq
Any consistent solution to this equation requires $w=1/3$
as it should be since the universe is to be dominated by radiation,
and the necessary solution is given by
\beq
\label{e:erh3}
\rho_R(t) = {3\over 32\pi G}{1\over t^2}
-\rho_\phi(t_i)\,{t_i^2\over t^2}\,\e^{-(w_\phi+1)\Delta_\phi(t-t_i)}.
\eeq
The first term 
leads to the maximum temperature 
for $t_i\sim M_{\rm pl}/\sqrt{\rho_\phi(t_i)}$, which is given by
\beq
\label{e:mt}
T_{\rm max}\sim \rho_\phi(t_i)^{1/4}.
\eeq
For the reheating temperature, from eq.(\ref{e:ev2}) we obtain
\beq
\label{e:rhta}
t\sim {1\over(w_\phi+1)\Delta_\phi}
\eeq
such that 
\beq
\label{e:rht}
T_{\rm RH}\sim \sqrt{w_\phi+1}\sqrt{M_{\rm pl}\Delta_\phi},
\eeq
which is higher than the reheating temperature for 
$w_\phi=0$\cite{KT,arXiv:1001.2600}. 
Note that $t\sim 1/\Delta_\phi$ is not proper in our model 
because it makes the reheating period rather too long
for small $\Delta_\phi$. 
Thus eq.(\ref{e:rhta}) should be the correct one.
Indeed, this reheating temperature can become sufficiently high 
if $w_\phi\gg 1$ even for small $\Delta_\phi$,
hence it justifies the word ``ekpyrotic."

If $\Delta_\phi= 0$, the annihilation term can no longer be ignored, then
the solution to eq.(\ref{e:1}) is given by
\beq
{1\over\rho_\phi(t)} 
\simeq {1\over\rho_\phi(t_i)}\left({t\over t_i}\right)^{2\alpha}
+{\alpha(w+1)\Sigma_\phi\over 2\alpha -1}\,
t\left({t\over t_i}\right)^{2\alpha-1},
\eeq
where $\alpha\equiv (w_\phi+1)/(w+1)$ and $\Sigma_\phi$ is taken as the 
average value.
As $w_\phi\gg 1$, i.e. $\alpha \gg 1$, $\rho_\phi(t)$ decreases because 
$t/t_i >1$. Using eqs.(\ref{e:erh1})(\ref{e:erh2}) for $\alpha \gg 1$, 
in this case, we obtain
\bseq
\begin{align}
\rho_R&\simeq {3\over 8\pi G}H^2-\rho_\phi \\
&\simeq {1\over 6\pi G(w+1)^2}{1\over t^2} 
- \left({t_i\over t}\right)^{2\alpha}
{\rho_\phi(t_i)\over 1 +\half(w+1)\Sigma_\phi t_i \rho_\phi(t_i)}.
\end{align}
\eseq
For $t_i\sim M_{\rm pl}/\sqrt{\rho_\phi(t_i)}$,
the maximum temperature is eq.(\ref{e:mt}) as before.
With
\beq
{1\over t}\sim \Sigma_\phi\rho_\phi(t_i)
\sim n_\phi(t_i)\langle\sigma v\rangle,
\eeq
the reheating temperature is
\beq
\label{e:rht2}
T_{\rm RH}\sim \left(M_{\rm pl}{1\over t}\right)^{1/2}
\sim \sqrt{M_{\rm pl}n_\phi(t_i)\langle\sigma v\rangle},
\eeq
which can be sufficiently high since $n_\phi(t_i)\langle\sigma v\rangle$
is large, particularly, for massless inflaton.
Note that this is true only if $w_\phi\gg 1$, hence ekpyrotic.

The reheating temperature can be constrained by 
CMB\cite{Kawasaki:2000en,Martin:2010kz,Mielczarek:2010ag}.
Ref.\cite{Mielczarek:2010ag} claims $T_{\rm RH}\sim 10^6$ GeV.
In $\Gamma_\phi\neq 0$ case, this can be achieved with Yukawa coupling constant 
$\lambda_f\sim 10^{-4}$ for massive inflaton with $m_\phi\sim 1$ TeV
(see eq.(\ref{e:pdr})) even for $w_\phi=0$. 
For smaller Yukawa couplings, sufficiently large $w_\phi$ can achieve
 the desired reheating temperature even for massless inflaton.
This indicates that the reheating temperature
given by eq.(\ref{e:rht}) (or eq.(\ref{e:rht2})) is actually sufficient 
and preheating is unnecessary in our model.
If $\Gamma_\phi > H_0\sim 10^{-26}\ {\rm sec}^{-1}$ such that the inflaton 
decays entirely during reheating, then $\phi$ cannot be DM.
However, for $w_\phi\gg 1$, with sufficiently small Yukawa coupling constants, 
we can achieve $\Gamma_\phi< H_0$ so that there is a possibility 
that the remaining inflaton can become DM after ekpyrotic reheating. 

\newsection{Inflaton as Dark Matter}


Let us now consider the case in which the remaining inflaton can become
DM after successful reheating. Once $\phi$ becomes massive, 
$\phi$ no longer behaves like a (uniform) fluid such that
$w_\phi=0$ is possible. 
In our model, the inflaton DM is basically Yukawa interacting scalar 
DM. (See \cite{Carone:2011iw} for another example.)
We assume that Yukawa couplings are flavor 
conserving so that we do not have to worry about $f\to f\phi$ process for 
massive $\phi$. Then, the relic density can be computed in terms of 
the usual Boltzmann eq. for annihilation/creation process\cite{Scherrer:1985zt}
\beq
\label{eP12}
\dot{n}_\phi +3Hn_\phi
=-\langle\sigma v\rangle \left(n^2_\phi - n_{\rm EQ}^2\right).
\eeq
Let $Y\equiv {n_\phi/s_E}$ and $x\equiv m_\phi/T$, where $s_E$ is the total 
entropy density of the universe and $T$ is the photon temperature, 
then, generalizing ref.\cite{Gondolo:1990dk} for arbitrary $w$, 
the Boltzmann eq. can be expressed as
\beq
\label{eP17}
{dY\over dx}+c_w{Y\over x}
=-\sqrt{{\pi\over 45 G}}{g_T^{1/2}m_\phi\over x^2}
\langle\sigma v\rangle\left(Y^2-Y^2_{\rm EQ}\right),
\eeq
where
\bseq
\begin{align}
\label{eP18}
g_T^{1/2} &\equiv 
{h_{\rm eff}\over\sqrt{g_{\rm eff}}}
\left(1+{1\over 3}{d\ln h_{\rm eff}\over d \ln T}\right),\\
\label{eP16b}
\rho&=g_{\rm eff}(T){\pi^2\over 30}T^4,\\
\label{eP16c}
s_E&=h_{\rm eff}(T)(1+w){\pi^2\over 30}T^3,
\end{align}
\eseq
and
\beq
c_w\equiv {4\over w+1}-3=
\begin{cases}
0, &\mbox{for radiation-dominated era,}\\
1,&\mbox{for matter-dominated era.}
\end{cases}
\eeq
Note that the prefactor $(1+w)$ in $s_E$ is due to $s_E\propto (\rho+p)$.

A relativistic derivation of $\langle\sigma v\rangle$
is given in \cite{Gondolo:1990dk}, which reads in the lab frame
\beq
\label{e:esv11}
\langle\sigma v\rangle=
{2x\over K_2^2(x)}\int_0^\infty d\epsilon\, 
\sigma v_{\rm lab}\sqrt{\epsilon}(1+2\epsilon)K_1(2x\sqrt{1+\epsilon}),
\eeq
where $\epsilon\equiv {s\over 4m^2} -1$ is the average kinetic energy 
per mass for the Mandelstam variable $s$, 
and $K_n(x)$ is the modified Bessel function of the second kind of order $n$.
Note that this formula works as long as two incident particles are collinear.
If $\phi$ is sufficiently massive as in our case, 
we can use non-relativistic expansion of 
$\sigma v_{\rm lab}$ in powers of $\epsilon$ as
\beq
\label{e:esv12}
\sigma v_{\rm lab} =\sum_{n=0} {a_n\over n!} \epsilon^n
\eeq
such that
\beq
\label{e:esv13}
\langle\sigma v\rangle=a_0 +{3\over 2}{a_1\over x}+\CO(x^{-2})
\equiv \sum_{n=0}{\langle\sigma v\rangle_n\over x^n}.
\eeq

For Yukawa coupling $\lambda_f\phi\bar{f}f$, 
the cross section for $\phi\phi\to\bar{f} f$ in the CE frame is given 
by\footnote{For the purpose of the dark matter relic density, effective Yukawa 
couplings between the inflaton and low energy fermions can be generated at 
higher orders after the EW symmetry breaking in terms of, e.g., the gauge 
interactions between isospin singlet fermions and other SM-like fermions.}
\beq
\label{esv15}
\sigma_{\phi\phi\to f\bar{f}}
={\lambda_f^4\over 16\pi}{p_f\over p_i}{1\over E_i^2}
\left|1+{(m_\phi^2-4m_f^2)^2\over m_\phi^4-4m_\phi^2 m_f^2+4m_f^2 E_i^2}
-{2E_i^2-(m_\phi^2-4m_f^2)\over 2 p_i p_f}
\ln\left({2E_i^2-m_\phi^2+2p_i p_f\over 2E_i^2-m_\phi^2-2p_i p_f}\right)
\right|,
\eeq
where $p_f\equiv (E_i^2-m_f^2)^{1/2}$. 
Expanding this according to eq.(\ref{e:esv12}), we can obtain
\beq
a_0 =a_{{\rm CE},0}= {\lambda_f^4\over 8\pi}
{\alpha_f^2(10-\alpha_f^2)(1-\alpha_f^2)^{1/2}\over m_\phi^2},
\eeq
where $\alpha_f\equiv m_f/m_\phi$. As $\alpha_f\to 0$, $a_0\to 0$ and
the next nonvanishing expansion coefficient is
\beq
a_1=a_{{\rm CE},1}-a_{{\rm CE},0}
= {\lambda_f^4\over 3\pi}{1\over  m_\phi^2}.
\eeq
If $\phi$ is massless, which is relevant for the reheating process, 
this cross section develops an infrared singularity,
which can be regularized by introducing a minimal cutoff initial energy.

With eq.(\ref{e:esv13}), eq.(\ref{eP17}) now reads, in the leading order of $x$,
\beq
\label{eP20}
{dY\over dx}+c_w{Y\over x} = -A_n{Y^2 -Y_{\rm EQ}^2\over x^{2+n}},
\eeq
where $Y_{\rm EQ}$ is to be suppressed by $\e^{-x}$, $n$ accounts 
the leading nonvanishing term, and
\beq
\label{eP21}
A_n\equiv 
{1\over 3}\sqrt{{\pi\over 5 G}}g_T^{1/2}m_\phi \langle\sigma v\rangle_n.
\eeq
Taking average value of $g_T^{1/2}$ for an approximation and
integrating from $x_F$ to $x_0$, i.e. from the freeze-out to the present, 
we can obtain a solution to this equation as
\beq
\label{eP23}
{x_0^{n+1}\over Y_0}\simeq \left({x_F^{n+1}\over Y_F}+{A_n\over 1+n+c_w}\right)
\left({x_0\over x_F}\right)^{1+n+c_w},
\eeq
where we assume $x_0/x_F=T_F/T_0\gg 1$.

If the density evolution overlaps from radiation-dominated to matter-dominated
era, we should solve eq.(\ref{eP20}) consecutively from $x_F\to x_c\to x_0$,
where $x_c$ is for the cross-over from radiation-dominated to matter-dominated era. Then, the current relic density can be better approximated as
\beq
\label{eP23e}
{x_0^{n+1}\over Y_0}
\simeq \left({x_F^{n+1}\over Y_F}+{A_n\over n+1}\right)
\left({x_0\over x_F}\right)^{n+1}
\left({x_0\over x_c}\right),
\eeq
where $x_0/x_c=T_c/T_0\sim 10^4$. 


If $\langle\sigma v\rangle$ is large enough, the freeze-out temperature 
can be determined by
\beq
\label{eP4b}
H(t_F) = n_{\rm EQ}\langle\sigma v\rangle,
\eeq
where, for large $x$,
\beq
\label{eP1c}
n_{\rm EQ} ={g_f T^3\over 2\pi^2} x^2K_2(x)
\simeq {g_f T^3\over 2\pi^2}\sqrt{\pi\over 2}x^{3/2}\e^{-x}.
\eeq
Then, assuming freeze-out takes place during the radiation-dominated era,
the freeze-out value $Y_F$ is given by
\beq
\label{eP29}
Y_F=Y_{\rm EQ}(x_F) = {n_{\rm EQ}(x_F)\over s_E(x_F)}
\simeq 
{15\over \sqrt{2\pi}}{\sqrt{g_{\rm eff}(T_F)}\over h_{\rm eff}(T_F)} 
{x_F^{1+n}\over M_{\rm pl}m_\phi \langle\sigma v\rangle_n}.
\eeq


From eq.(\ref{eP23}) and eq.(\ref{eP29}), assuming the density has evolved 
as if the entire period were radiation-dominated era, we can obtain
\beq
\label{eP30}
Y_0=B{x_F^{1+n}\over M_{\rm pl}m_\phi \langle\sigma v\rangle_n},
\eeq
where
\beq
B\equiv \left[{\sqrt{2\pi}\over 15}
{h_{\rm eff}(T_F)\over \sqrt{g_{\rm eff}(T_F)}}
+{1\over 3}\sqrt{{\pi\over 5}}{\sqrt{\bar{g_{\rm eff}}}\over 1+n}
\right]^{-1}.
\eeq
If we use eq.(\ref{eP23e}) to compensate the density evolution
during the matter-dominated era, 
\beq
\label{eP26p}
Y_0=B\left({x_c\over x_0}\right)
{x_F^{1+n}\over M_{\rm pl}m_\phi\langle\sigma v\rangle_n}.
\eeq
Note that $Y_0$ in this approximation is $10^{-4}$ times $Y_0$ of approximation
assuming only radiation-domination.
Then, with $h_{\rm eff}(T_0)=3.909$, $T_0=2.725\ {\rm K}$, and
$H_0=2.133\times 10^{-42}h\ {\rm GeV}$,
$\Omega_{\rm DM}$ is given by
\beq
\label{eP26n}
\Omega_{\rm DM}h^2=2.051\times 10^8 {m_\phi\over{\rm GeV}}Y_0.
\eeq
Comparing this to the current measured value 
$\Omega_{\rm DM}h^2=0.1126\pm 0.0036$\cite{Komatsu:2010fb}, 
we can obtain conditions on $m_\phi$ 
and $\lambda_f$. We will come back to this for a specific case later.

The above works only if the annihilation cross section is large enough
to satisfy eq.(\ref{eP4b}).
If it is too small so that eq.(\ref{eP4b}) cannot be satisfied at all, 
$n_{\rm EQ}(x_F)$ should be used to determine the relic density. 
The freeze-out parameter $x_F$ for $\langle\sigma v\rangle\simeq 0$
can be determined from the following equation\cite{Gondolo:1990dk}:
\beq
\label{eP29a}
0\simeq {K_1(x)\over K_2(x)}
-{1\over x}{d\ln h_{\rm eff}(T)\over d\ln T}.
\eeq
For large $x$, up to $x^{-2}$,
\beq
{K_1(x)\over K_2(x)}\simeq\left(1+{3\over 8x}\right)
\left(1+{15\over 8x}+{105\over 128x^2}\right)^{-1}
\simeq 1-{3\over 2x}+{345\over 128x^2},
\eeq
then eq.(\ref{eP29a}) leads to an equation for $x_F$
\beq
x^2-{3\over 2}x +{345\over 128}-\xi m_\phi=0,
\eeq
where 
\beq
\xi\equiv {d\ln h_{\rm eff}(T)\over dT}
\eeq
can be approximated by taking a suitable numerical value.
With eqs.(\ref{eP23})(\ref{eP1c}) and $c_w=0$, 
\beq
\label{eP33b}
\Omega_{\rm DM}h^2 ={8\pi G h^2\over 3H_0^2}m_\phi n_\phi(t_0)
=7.814\times 10^{44} {m_\phi T_0^3 \over ({\rm GeV})^4}
{h_{\rm eff}(T_0)\over h_{\rm eff}(T_F)} x_F^{3/2}\e^{-x_F},
\eeq
where $T_0=2.35\times 10^{-13}\ {\rm GeV}$.
Using $h_{\rm eff}(T_0)=3.909$ and $h_{\rm eff}(T_F)\simeq 10$,
$\xi=0.83\ {\rm GeV}^{-1}$ and $x_F\simeq 30$ for  $m_\phi\simeq 1\ {\rm TeV}$.  


Note that $\phi\to \bar{f} f$ requires $m_\phi > 2m_f$, i.e.
$1/2>\alpha_f$, while $\phi\phi\to \bar{f} f$ requires $m_\phi > m_f$
i.e. $1>\alpha_f$. So, there are three possibilities:
\bseq
\begin{align}
\alpha_f \geq 1&:\quad \sigma=0,\ \Gamma_\phi=0, \\
1>\alpha_f \geq \half&:\quad \sigma\neq 0,\ \Gamma_\phi=0, \\
\half > \alpha_f &:\quad \sigma\neq 0,\ \Gamma_\phi\neq 0.
\end{align}
\eseq
Since we expect $m_\phi \gtrsim 1\ {\rm TeV}$, $\alpha_f \geq 1$ for all 
flavors cannot happen. So, the fate of the inflaton depends on these mass ranges 
as well as Yukawa coupling constants and there are three different cases:

$\bullet$ Case I: The inflaton is not a DM candidate and decays entirely 
before present day.

$\bullet$ Case II: The inflaton is a DM candidate and the relic density is 
given by eq.(\ref{eP33b}).

$\bullet$ Case III: The inflaton is a DM candidate and the relic density 
is given by eq.(\ref{eP26n}).

If there is at least one light fermion flavor with 
$\lambda_f \gtrsim 10^{-26}$ 
(i.e. $\Gamma_\phi\gtrsim 10^{-26}\ {\rm sec}^{-1}$), 
it is Case I. If $\lambda_f \lesssim  10^{-26}$ for all flavors, it is Case II.
Case III is the case with some flavors with $1>\alpha_f \geq {1\over 2}$
and other flavors with ${1\over 2} > \alpha_f$. A good example of Case III
is the technicolor (TC) model with Yukawa interacting sterile 
scalars\cite{Kephart:2011wu}, in which technifermions have 
$\lambda_{\rm TC} > 10^{-26}$
while quarks and leptons have $\lambda_{Q,L} \lesssim 10^{-26}$.
Another possibility of Case III without TC is to allow the fourth generation
of suitable mass with $\lambda_4 >10^{-26}$.

Let us look into the TC case more in detail.
For $m_\phi\simeq 1\ {\rm TeV}$, at least the lightest technifermion 
mass should be between $1\ {\rm TeV}$ and $500\ {\rm GeV}$. Then, the density
evolution is dominated by the annihilation process so that we can use
eq.(\ref{eP26p}) to determine the relic density. For example, if we assume
there is only one technifermion within the mass range with mass about
600 GeV, then $\lambda_f\simeq 3.4\times 10^{-2}$ leads to 
$m_\phi\simeq 1\ {\rm TeV}$ with $x_F\simeq 18$. 
If we use eq.(\ref{eP30}) just for radiation-dominated
era, we get $\lambda_f\simeq 0.38$ with $x_F\simeq 28$. Since the difference
is significant enough, we believe that the relic density is better approximated
by eq.(\ref{eP26p}). This is because a significant part of relic density 
evolution takes place during the matter-dominated era so that it should not 
be ignored.
If we have more technifermions within the mass range, smaller Yukawa coupling
constants can lead to $m_\phi\simeq 1\ {\rm TeV}$.

\newsection{Discussions}

In this paper, we have demonstrated that a successful cosmological model
could be in principle achieved with just two sterile scalar fields. 
In cases II and III,
the remainder of the inflaton after reheating can become a DM candidate.
In this context, it may be possible that massive inflaton can trigger
local clustering of matter even during the radiation-dominated era,
hence it can become an early seed of structure formation.
The other sterile scalar field as DEF behaves more like a free field and
is uniformly distributed with its flat potential, 
whose magnitude is the same as the absolute value of 
the inflaton potential energy at a new vacuum after inflation,
so that the inflaton's vacuum fluctuation after reheating can account 
for the magnitude of dark energy. We leave the computation of the amount of 
this vacuum fluctuation as a future task. Of course, we can make $\eta$
more dynamical by introducing a nontrivial potential energy and/or by 
adding sufficiently weak interactions to the inflaton, but due to the
observational constraint indicating no dynamics\cite{Ade:2013ktc}, 
we think it is a reasonable assumption at this moment. 
However, it will be interesting to work out if any observable dynamics out 
of $\eta$, which we will also leave as a future work.

It will be also interesting to solve eq.(8) for time-dependent $w_\phi$ 
for comparison. We leave this also as a future work.

The inflaton potential we have used has an analogous form to the Fermi
surface so that it could indicate a more fundamental origin of the potential.
This could also explain proper scale for parameters introduced in the potential
so that we can describe more detailed behavior of the inflaton.  
Also, even though we have not used the oscillatory nature of the inflaton 
field explicitly (eq.(\ref{e:1}) is consistent with implicit usage of
it\cite{KT}),
we suspect there might be a connection between our ekpyrotic reheating
and the fermionic preheating considered in ref.\cite{Greene:2000ew} 
to a certain level. It will be interesting to see if this is indeed the case.

\noindent
{\bf Acknowledgments:}

I would like to thank Bob Scherrer for many informative conversations, 
reading the manuscript and helpful comments.

\renewcommand{\Large}{\large}

\end{document}